\documentclass[10pt,conference,letterpaper]{IEEEtran}
\usepackage{times,amsmath,epsfig}
\usepackage{epsfig}
\usepackage{epstopdf}
\usepackage{amsmath}
\usepackage{amsfonts}
\usepackage{subfigure}
\usepackage[ruled,vlined, linesnumbered]{algorithm2e}
\title{CrowdPlanner: A Crowd-Based Route Recommendation System}
\author{%
{Han Su }%
\vspace{1.6mm}\\
\fontsize{10}{10}\selectfont\itshape
University of Queensland, Australia \\
\fontsize{9}{9}\selectfont\ttfamily\upshape
h.su1@uq.edu.au
}

\newtheorem{definition}{Definition}

\begin{document}
\maketitle

\begin{abstract}
As travel is taking more significant part in our life, route recommendation service becomes a big business and attracts many major players in IT industry. Given a user specified origin and destination, a route recommendation service aims to provide users the   routes  with the best travelling experience according to criteria such as travelling distance, travelling time, traffic condition, etc. However, previous research shows that even the routes recommended by the big-thumb service providers can deviate significantly from the routes travelled by experienced drivers. It means travellers' preferences on route selection are influenced by many latent and dynamic factors that are hard to be modelled exactly with pre-defined formulas.  In this work we approach this challenging problem with a completely different perspective -- leveraging crowds' knowledge to improve the recommendation quality. In this light, CrowdPlanner -- a novel crowd-based route recommendation system has been developed, which requests human workers to evaluate candidates routes recommended by different sources and methods, and determine the best route based on the feedbacks of these workers. Our system addresses two critical issues in its core components: a) task generation component generates a series of informative and concise questions with optimized ordering for a given candidate route set so that workers feel comfortable and easy to answer; and b) worker selection component utilizes a set of selection criteria and an efficient algorithm to find the most eligible workers to answer the questions with high accuracy. 

\end{abstract}


\section{Introduction}

Travelling plays a vital role in our daily life. 
Thanks to  the rapid development of GPS technologies and flourish of navigation service providers (e.g., Google Map, Bing Map, TomTom), we can now travel to unfamiliar places with much less effort, by simply following the recommended routes. While the detailed mechanisms that are adopted to recommend routes are different, travelling distance and time are the most important criteria and factors in those recommendation algorithms, which results in the shortest route and/or fastest route. With increasing number of users who rely on these map services to travel, a natural question arises: {\em are these routes always good enough to be the best choice when people travel?} Ceikute et al~\cite{Ceikut2013Routing} are the first to assess the routing service quality by comparing the popular routes, the ones most drivers prefer, and the  routes recommended by a big thumb map service provider. Their results show that the there are big distances between popular routes and recommended routes. It concludes that experienced/frequent drivers' preferences do not always agree with the routes recommended by navigation service. Actually the cause of this phenomenon is not difficult to find out: drivers preferences are influenced by lots of factors in addition to distance and time, such as the number of traffic lights, speed limitation, road condition, weather, amongst many others, which are very difficult to be taken into consideration simultaneously by a single routing algorithm. That is to say, driver's preference is the ultimate criterion to judge the goodness of a route, i.e., given a source and a destination, route $A$ is regarded as a better choice than route $B$ if more drivers prefer to drive along  $A$ for some reasons. 

{\em Then can we find the most popular routes between two places and return them to the users?} Ideally the routes recommended in this way should have improved quality since the popular routes are usually obtained by mining historical trajectories of drivers and can better  indicate the driver's preference. But after further investigation, we have observed some problems of this approach. First, since users can specify two arbitrary places as source and destination, it happens that we cannot get enough support from historical trajectory data to obtain reliable results. Even worse, in areas where historical data are very sparse, the ``popular'' route can refer to a bad one with some random historical data. Therefore, a system that solely relies on those popular routes can perform badly at large scale. Second, there exist a number of popular route mining algorithms, each claiming some superiority from certain perspectives. When the results of these algorithms disagree with each other, it is still a pain for strangers to make a wise choice.

%

As we can see from the above analysis, it is not uncommon that routes recommended by different ways disagree with each other. This makes choosing the best route to be a very subjective question, which is hard for a computer algorithm to answer. Inspired by the emerging concept of crowd sourcing that explicitly leverages human knowledge to resolve complex problems, we propose a novel crowd-based route recommendation system -- CrowdPlanner, which combines the strengths from computers and human brains to recommend the best route with respect to the knowledge of experienced drivers. Instead of proposing new or optimizing existing routing algorithms, our work takes an entirely different approach by consolidating candidate routes from different sources (e.g., map service providers, popular routes) and requesting experienced drivers to select amongst them. Our system will return the most promising one according to the selection of drivers.

It is a non-trivial task to perform route evaluation. 
First, it is not easy to automatically publish a user friendly task.  
Since the human's knowledge of routes are quite different from computers, that is human remember routes discretely while computer process routes continuously. 
So the system should intelligently select some typical points of the comparing routes instead of giving the continuous routes for surveyors to choose from.
Second, we should select the workers to ensure the quality of the task result since the quality of route evaluation is largely relying on the traveling experience of the workers.

In this paper, we propose a {\em CrowdPlanner system} for recommending the evaluated best route.
It comprises two layers: a mobile client supporting users to send a route recommendation request and answering surveys; 
a server which possesses the request by generating recommendation route candidates, evaluating candidate routes itself using existing evaluated routes and publish a route recommendation task. 
Especially in publishing a route recommendation task, a efficient question selecting and ordering method is proposed, which selects a set of significant landmarks to help workers select the their most recommend route;
an effective worker selecting method is also proposed in order to assign each task to its most familiar and eligible workers. 

To sum up, we make the following major contributions. 
\begin{itemize}
\item We make a key observation that the cognition of routes of humans is quite different from computers, that humans recognize routes by concrete points while computers recognize routes by continuous lines. 
\item We recommend the verified best route between two places to users, while the routes recommended by route recommendation web services are different from humans' preference and the different popular route detection algorithms provide different popular routes. 
\item We conduct extensive experiments based on hundreds of volunteers and  large-scale real
trajectory dataset, which empirically demonstrates that the
CrowdPlanner system can always provide best route between places efficiently and the survey mechanism is intelligent and user friendly. 
\end{itemize}

\section{Problem Statement}\label{sec:problem}
In this section, we present some preliminary concepts and give an overview of the CrowdPlanner system. Table~\ref{tb:notation} summarized the major notations used in the rest of the paper.

\begin{table}[htb]

\caption{Summarize of notations}\label{tb:notation}
\centering\small

\begin{tabular}{|l|l|}
\hline 
{\bf \centering Notation} & {\bf \centering Definition}\tabularnewline
\hline 
$R$ & a recommended route \tabularnewline
\hline 
$\mathbb{R} $ & candidate set of recommended routes\tabularnewline
\hline 
$p$ & a place in the space\tabularnewline
\hline 
$l$ & a landmark in the space\tabularnewline
\hline 
$l.s$ & significance of landmark $l$ \tabularnewline
\hline 
$\mathbb{L}$ & a landmarks set\tabularnewline
\hline 
$\mathbb{L}_{\mathbb{R}} $ & the questioned landmark set of routes set $\mathbb{R}$ \tabularnewline
\hline 
$d(l_i,l_j)$ & Euclidean distance between landmarks $l_i$ and $l_j$\tabularnewline
\hline 
$w$ & a worker of the system\tabularnewline
\hline 
$\mathbb{W}_{\mathbb{R}}$& the selected workers of routes set $\mathbb{R}$\tabularnewline
\hline 
\end{tabular}
\end{table}

\subsection{Preliminary Concepts}

\begin{definition}{Route:} 
A route $R$ is a continuous travelling path. 
We use a sequence $[p_1, p_2, \cdots, p_n]$, which consists of a source, a destination, and a sequence of consecutive road intersections in-between, to represent a route. 
\end{definition}

\begin{definition}{Landmark:}
A landmark is a geographical object in the space, which is stable and independent of the recommended routes. A landmark can be either a point (i.e., Point Of Interest), a line (i.e., street and high way) or a region (i.e., block and suburb) in the space.
\end{definition}

\begin{definition}{Landmark-based Route:}
A landmark-based route $\bar R$ is a route represented as a finite sequence of landmarks, i.e., $\bar R = \{l_1,l_2,...,l_n\}$.
\end{definition}

In order to obtain the landmark-based route from a raw route, we employ our previous research results on anchor-based trajectory calibration~\cite{su2013calibrating} to rewrite the continuous recommend routes into landmark-based routes, by treating landmarks as anchor points.

\begin{definition}{Discriminative landmarks:}
A landmark set $\mathbb{L}$ is called {\em discriminative} to a set of landmark-based routes $\bar {\mathbb{R}}$ if for any two routes $\bar R_1$ and $\bar R_2$ of $\bar {\mathbb{R}}$, the joint sets $\bar {R_1} \cap \mathbb{L}$ and $\bar  {R_2}\cap \mathbb{L}$ are different.
\end{definition}

For example, $\mathbb{L}_1 =\{l_3, l_4\}$ is discriminative to $ {R_1}=\{l_1, l_2, l_3\}$ and ${R_2}=\{l_1, l_2, l_4\}$, since the joint sets  $ {R_1}\cap \mathbb{L}_1=\{l_3\}$ and $ {R_2}\cap \mathbb{L}_1=\{l_4\}$ are different, but $\mathbb{L}_2 =\{l_1, l_2\}$ is not discriminative to $R_1$ and $R_2$. 

\begin{definition}{Simplest Discriminative:}
An identifiable set $\mathbb{L}$ to $\bar {\mathbb{R}}$ is simplest discriminative to $\bar {\mathbb{R}}$ if removing any landmark from $\mathbb{L}$, $\mathbb{L}$ is not discriminative to $\bar {\mathbb{R}}$ any more.
\label{define:SimplestIdentifiable}
\end{definition}
Continuing with the previous example, $\mathbb{L}_1$ is not simplest identifiable to $\bar {\mathbb{R}}$, since after removing $l_3$ from $\mathbb{L}_1$ it still identifiable to $\bar {\mathbb{R}}$, while sets $\mathbb{L}_3=\{l_3\}$ and $\mathbb{L}_4=\{l_4\}$ are. 




\subsection{Overview of CrowdPlanner}

\begin{figure}
\centering
	{\psfig{figure=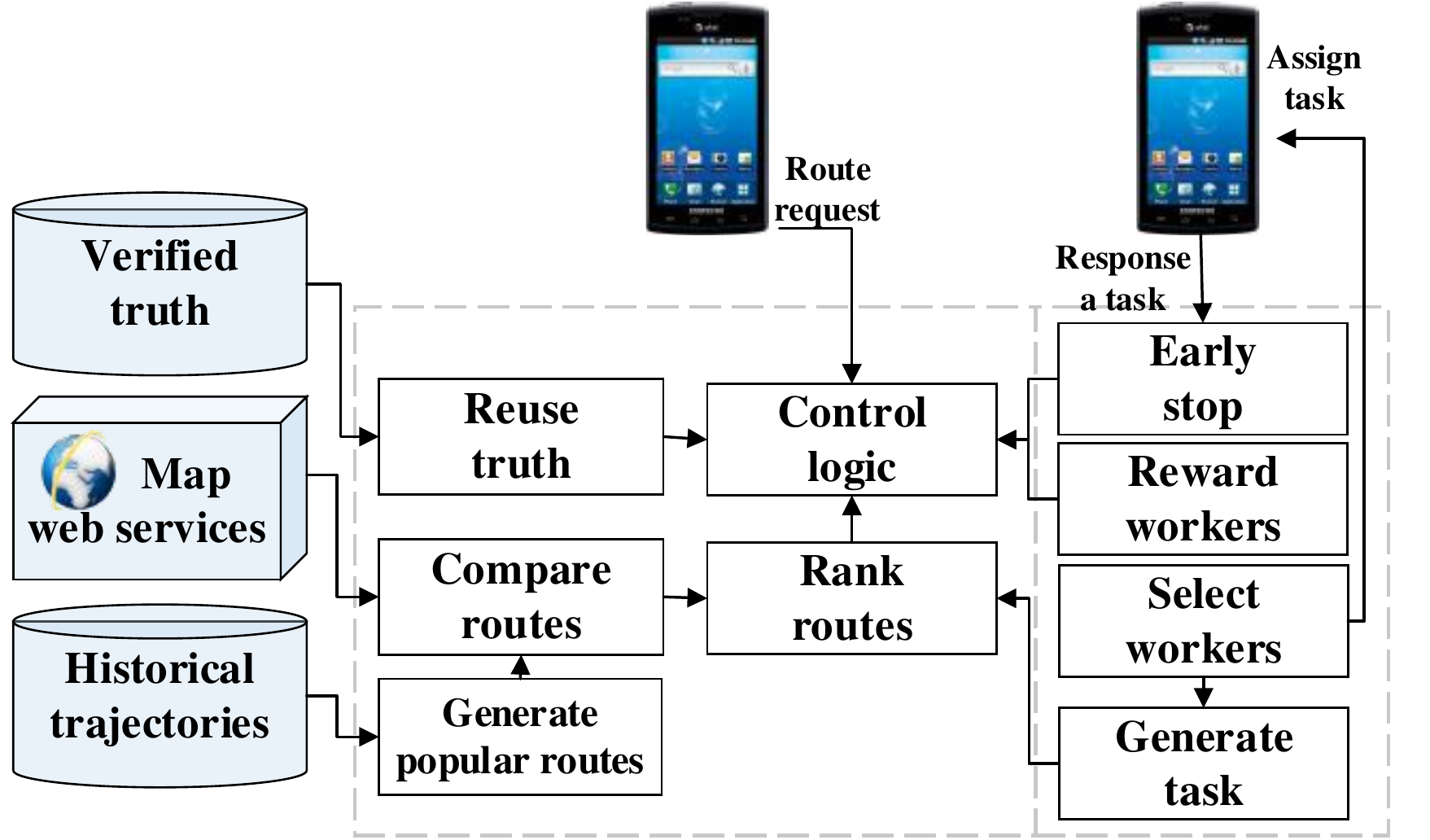,width=80mm} }
	\caption{System Overview}
	\label{fig:structure}
\end{figure}

CrowdPlanner is a two-layer system (mobile client layer and server layer) which receives user's request from mobile client specifying the source and destination, processes the request on the server and finally returns the verified best routes to the user. 
Fig.~\ref{fig:structure} shows the overview of the proposed CrowdPlanner system, which comprises two modules: traditional route recommendation (TR) and crowd-based route recommendation (CR).
The workflow of CrowdPlanner is as follows: the TR module firstly processes user's request by trying to evaluating the quality of candidate routes obtained from external sources such as map services and historical trajectory mining;
the CR module will generate a crowdsourcing task when the TR module can not distinguish the quality of candidate routes, and return the best route based on the feedbacks of human workers of the system. 

\subsubsection{Traditional Route Recommendation Module}

This module processes the user's request by generating a set of candidate routes from external sources (route generation component) and evaluating the quality of those routes automatically without involving human effort (route evaluation component). 

{\bf Control logic component:} 
This component receives the user's request and controls the workflow of the entire system. It also coordinates the interactions between the TR module and CR module. 
Once a user's request is received by the control logic component, it will invoke {\em reuse truth component} to match the request to the verified routes (truth) between two places at his departure time. 
If the new coming request is a hit of the truth, the system will return result immediately.
Otherwise the component will invoke the {\em route generation component} to automatically generate some candidate routes and {\em route evaluation component} to evaluate the qualities of these candidate routes using the verified truth.


{\bf Route evaluation component:} 
This component evaluates the routes using computer power and it provides an efficient way to reduce the cost of CrowdPlanner, since it can largely reduce the amount of tasks generated.
The component will firstly build up a candidate route set by invoking {\em route generation component}. 
If some of these routes agree with each other to a high degree, one of them will be selected as the best recommended route and added into a truth database with the corresponding time tag. 
If a best recommended route can not be determined, the system will assign each candidate route a confidence score, which is generated by the verified truths and illustrates the possibility of the route to be the best recommended route.
A route with the highest confidence score that is greater than a threshold $\eta$ will be regarded to be the best recommended and returned to the user; 
otherwise the logic control will hand over the request to the CR module.


{\bf Route generation component:} This component generates two types of candidate routes, the one provide by web services such as Google Map and the one generated from historical trajectories by using popular route mining algorithms, i.e., MPR, LDR and MFP.


\subsubsection{Crowd Route Recommendation Module}
Crowd route recommendation module will take over the route recommendation request when the traditional route recommendation module cannot provide the best route with high enough confidence.
The module will generate a crowdsourcing task consisting of a series simple but informative binary questions (task generation component), and assign the task to a set of selected worker who are most suitable to answer these questions (worker selection component).

{\bf Task generation component:} 
As the core of CrowdPlanner, this component generates a task by proposing a series of questions for workers to answer. It is beneficial to have these questions as simple and compact as possible, since both the accuracy and economic effectiveness of the system can be improved. The design of this component will address two important issues: {\em what to ask in questions} and {\em how to ask the questions}. We will discuss the detailed mechanism of this part in Section~\ref{sec:generatequestion}.

{\bf Worker selection component:} 
This is another core component of CrowdPlanner. 
In order to maximize the effectiveness of the system, we need to select a set of eligible workers who are most suitable to answer the questions in a given task, by estimating the worker's familiarity with the area of request. Technical details of this component will be presented in Section~\ref{sec:targetworker}. 

{\bf Early stop component:} 
In most cases, we do not to need to wait for all the answers of the assigned workers.
When partial feedbacks have been collected, this component will evaluate the confidence of the answer and return the result to the user as early as possible when the confidence is high enough. 

{\bf Rewarding component:}
This component rewards the workers according to their workload and the quality of their answers. The reward points can be used later when they request a route recommendation in CrowdPlanner. 

\vspace{1em}
In the following two sections, we will present the design and technical details of the two core components of CrowdPlanner: task generation and worker selection. 

\section{Task Generation} \label{sec:generatequestion}

\begin{figure}[h]
\centering
	{\psfig{figure=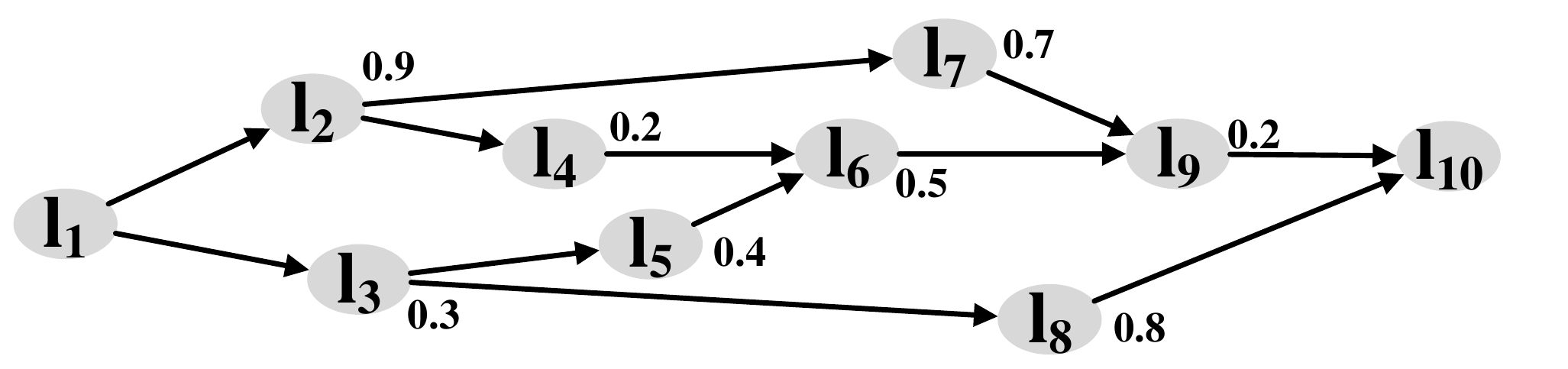,width=75mm} }
	\caption{An example of landmark-based recommended routes between $l_1$ and $l_{10}$}
	\label{fig:routes}
\end{figure}

Almost everyone has the experience that you can not explain a route clearly to someone, even you know exactly how to get there in mind, 
which implies  describing a route concretely is hard for human. 
Therefore we can not simply publish a task to workers that requires them to describe the best route in a turn-by-turn manner. 
As an alternative and more friendly way, we may provide several pictures, which demonstrate these candidate routes on a map, as multiple choices for workers to choose.
Taken the route recommendation request in Fig.~\ref{fig:routes} as example, we publish a multiple choice question to workers by showing four routes on a map and ask them to pick the route they most prefer. 
Although all the routes have been visualized on a map, it still requires lots of efforts to realize the differences among all candidate routes, especially when they use smartphones with small screens to do this job. 
We observe that  humans like to use a set of significant locations, i.e., landmarks, to describe a route in high level rather than a sequence of continuous roads as computers do. So to make the question easier to answer, we  summarize the candidate routes using landmarks and present the differences to workers, instead of asking them to find out by themselves. 
Besides, how the questions are presented can also affect the complexity of a task. 
For example, a multiple choice question with all candidate routes is more difficult to answer than a couple of binary questions such as `` do you prefer the route passing landmark $A$ at 2:00pm?'', since~\cite{su2012crowdsourcing} pointed out that several binary choice questions are easier and more accurate than a multiple choice question. 
Based on the above analysis, we will generate a task as a sequence of binary question, each relating to a landmark that can discriminate some of the candidate routes. 

In this section, we will present in detail our task generation process, which can be divided into three phases: inferring landmark significance, landmark  selection and question ordering.
Specifically, the first phase infers the significance for each landmark which indicates people's familiarity with it. 
The second phase tries to use a set of most significant landmarks to summarize the difference among the candidate routes.
The third phase generates the final task by ordering the questions in a smart way so that the expected number of issued questions is as small as possible. 

\subsection{Inferring Landmark Significance}

It is a common sense that landmarks has different significances. For instance, the White House is world famous, but Pennsylvania Ave, where the White House is located on, is only known by some locals of Washington DC.
People tend to be more familiar with the landmarks that are frequently referred to by different sources, e.g., public praise, news, bus stop, yellow pages. In this work, we utilize the online check-in record in a popular location-based social network (LBSN) and trajectories of taxicabs in the target city to infer the significance of landmarks, since these two datasets are large scale enough to cover most areas of a city. By regarding the travellers as authorities, landmarks as hubs, and check-ins/visits as hyperlinks, we can leverage a HITS-like algorithm such as~\cite{zheng2009mining} to infer the significance of a landmark. Readers who are interested in the technical details can refer to~\cite{zheng2009mining}. 

\subsection{Landmark Selection}

Although any landmark can be used to generate a question, not all of them are suitable for the purpose of generating easy questions given a certain candidate route set $\bar{\mathbb{R}}$ (notably in this section we rewrite all the routes in $\mathbb{R}$ into landmark-based routes so we use $\bar{\mathbb{R}}$ to denote the candidate route set).
First, the selected landmark set $\mathbb{L}$ should be discriminative to the candidate routes $\bar {\mathbb{R}}$, which ensures that the difference between any two routes can be presented. 
Second, the landmarks of $\mathbb{L}$ should have high significance, so that more  people can answer the question accurately. 
Third, in order to reduce the work load of workers, the selected landmark set $\mathbb{L}$ should be as small as possible.
Therefore the problem of landmark selection is to find a small set of highly significant landmarks which are discriminative to all the candidate routes.
It can be formally represented as an optimization problem in below. \\
{\bf Given} $n$ landmark-based candidate routes $\bar {\mathbb{R}}$, and the significance of each landmark\\
{\bf Select} a landmark set $\mathbb{L}$ with the size of $k$ ($\left \lceil \log_2 (n)\right \rceil \le k \le n$) which is discriminative to $\bar{\mathbb{R}}$
\\
{\bf Maximize} $\sum\nolimits_{l \in \mathbb{L}} {l.s}\cdot |\mathbb{L}|^{-1}$\\
Here, the target function aims to maximize the product of two opposite factors: $\sum\nolimits_{l \in \mathbb{L}} {l.s}$, which is proportional to the sum of significances of $\mathbb{L}$ and $ |\mathbb{L}|^{-1}$, which is inversely proportional to the size of $\mathbb{L}$.

Due to the trade-offs between maximizing accumulate significance of the selected landmark set $\mathbb{L}$ and minimizing the size of $\mathbb{L}$ as well as  the restriction that $\mathbb{L}$ must be discriminative to $\bar{\mathbb{R}}$,  it is non-trivial to generate $\mathbb{L}$.
A straightforward method is to enumerate all the landmark combinations of $\bigcup\limits_{\bar R\in \bar{\mathbb{R}}}{\bar R}$ and find a discriminative landmark set with the maximized target value. 
However, the time cost grows exponentially with the size of landmark set, rendering this method impractical. 
To speed up this process,  we propose two landmark selecting algorithms, called Incremental Landmark Selecting and GreedySelect.  

\subsubsection{Incremental Landmark Selecting}

The main idea of ILS is to select several {\em simplest discriminative landmark} sets, each of which has different size (from $\left \lceil \log_2 n \right \rceil$ to $n$). 
The selected simplest discriminative set  of size $k$ 
has the maximum objective equation value among all the simplest discriminative landmark sets of size $k$. 
And then ILS can obtain the best landmark set using the selected simplest discriminative landmark sets.

At the very beginning, we need to roughly filter out some non-benificial landmarks which are on/ not on every candidate route. 
The beneficial landmarks set of $\bar {\mathbb{R}}$ can be generated as following:
$\mathbb{L} = \bigcup\limits_{\bar R \in \bar {\mathbb{R}}} {\bar R}  - \bigcap\limits_{\bar R \in \bar {\mathbb{R}}} { \bar R} $.
Because of optimization requirements, we sort $\mathbb{L}$ in descending order of their significances.
Notably, the sorted landmarks should be not stored in a set, but for convenience we still use $\mathbb{L}$ to denote to sorted beneficial landmarks.
We denote the selected simplest discriminative set of size $k$ by $\mathbb{L}_{sim[k]}$. 
The aim of the step is to find best landmark set $\mathbb{L}_{\bar {\mathbb{R}}}$ using $\mathbb{L}_{sim[k]}$, where $k$ ranges from $\left \lceil \log_2 n \right \rceil$ to $n$.
The discriminative set $\mathbb{L}_k$ of length of $k$ with maximum objective equation value can be calculated using $\mathbb{L}_{sim[i]}$ ($\left \lceil \log_2 n \right \rceil \le i \le k $) as following:
\begin{equation*}
\mathbb{L}_k = \arg \mathop {\max }\limits_{\left \lceil \log_2 n \right \rceil \le i \le k}{GetValue(GetMaxSet({\mathbb{L}}, k, \mathbb{L}_{sim[i]}))}
\end{equation*}
where the $GetMaxSet({\mathbb{L}}, k, \mathbb{L}_{sim[i]})$ is a function which returns the superset of $\mathbb{L}_{sim[i]}$ with the maximum objective equation value among all the $\mathbb{L}_{sim[i]}$'s supersets of size $k$ and 
$GetValue(\cdot)$ is a function which returns the objective equation value of $\cdot$.
$\mathbb{L}_k$ must be discriminative to $\bar {\mathbb{R}}$. 
Since the $\mathbb{L}$ is sorted, $GetMaxSet({\mathbb{L}}, k, \mathbb{L}_{sim[i]})$ has the time complexity of $O(1)$. 
Afterward, $\mathbb{L}_{\mathbb{R}}$ can be  generated as following:
\begin{equation*}
\mathbb{L}_{\mathbb{R}} = \arg \mathop {\max }\limits_{\left \lceil \log_2 n \right \rceil \le k \le n}{GetValue(\mathbb{L}_k)}
\end{equation*}
In the sequel, we will discuss how to obtain $\mathbb{L}_{sim[i]}$.

In order to obtain all the $\mathbb{L}_{sim[i]}$, we use a ``bottom up" approach, of which landmark sets are extended from size one to size $n$ by adding one landmark at a time and groups of landmark sets are tested.
Let $\mathbb{S}_{k}$ denote a set forming by several landmark combinations with the size of $k$.
The $\mathbb{L}_{sim[k]}$ generating can mainly divided into three steps.
(1) Start step: the algorithm start from $k= 1$, thus $\mathbb{S}_{1}$ stores all the combinations of each single landmark  of size $1$.
(2) Calculating step: then the algorithm tests whether each combination $S \in \mathbb{S}_{k}$ is discriminative.
Among all the discriminative sets of $\mathbb{S}_k$, the set with the maximum objective equation value assigns to $\mathbb{L}_{sim[k]}$.
Afterward all the discriminative combinations of $\mathbb{S}_{k}$ and their superset are pruned and removed from $\mathbb{S}_{k}$, since their superset are all discriminative  and their maximum objective equation value can be easily generated from $\mathbb{L}_{sim[k]}$. 
(3) Expansion step: If $k$ equals to $n$, the algorithm stops.
Otherwise we generates $\mathbb{S}_{k+1}$ from $\mathbb{S}_{k}$, which consists of all the undiscriminative landmark sets of size $k$.
Adding a new landmark $l$ ($l \in \mathbb{L}-S$) to $S$ ($S \in \mathbb{S}_k$) will form a element $S'$ of $\mathbb{S}_{k+1}$. 
However, the expansion cause same sets being generated several times, i.e., $\{l_1\}$ expends $l_2$ to $\{l_1, l_2\}$ and $\{l_2\}$ expends $l_1$ to $\{l_1, l_2\}$ too.
So we restrict the significance of adding landmark $l$ must be less than the significance of any landmark of $S$.
Thus $\mathbb{S}_{k+1}$ can be generated as following:
\begin{equation*}
\mathbb{S}_{k+1}=\{S'|S'=S \cup \{l\} \wedge l.s \le \min\limits_{l'\in S}{l'.s} \wedge l\in \mathbb{L} -S \wedge S\in \mathbb{S}_k\}
\end{equation*}
and repeat step 2.

\subsubsection{GreedySelecting}

The main idea of GreedySelecting is to incrementally adding a landmark with the highest significance among the possible landmarks to an discriminative landmark set and use some tight upper bounds to prune many landmark sets with low significance values.    
Let $S$ denote the current testing landmark set and
$\mathbb{L}_{best}$  the best landmark set which is discriminative and has the highest target value.
The landmark selection process can be divided into three steps:

{\bf Preparation step:} at the very beginning, we need to roughly filter out some non-beneficial landmarks, the ones cannot identify any routes of $\bar{\mathbb{R}}$. 
A straightforward way is to filter out landmarks which are on/ not on every candidate route. 
Thus the beneficial landmarks set of $\bar {\mathbb{R}}$ can be generated as following:
$\mathbb{L} = \bigcup\limits_{\bar R \in \bar {\mathbb{R}}} {\bar R}  - \bigcap\limits_{\bar R \in \bar {\mathbb{R}}} { \bar R} $.
Because of optimization requirements, we sort $\mathbb{L}$ in descending order of their significances.
Please note the sorted landmarks should be not stored in a set, but for convenience we still use $\mathbb{L}$ to denote to sorted beneficial landmarks.

{\bf Expansion step:} this step generates the test landmark set $S$. 
We recursively generate the test landmark set $S$, as shown in Algorithm~\ref{algo:expansion}.
In this step, we find all the landmarks not in $S$,
pick non-added biggest landmark of them, and add it to $S$.
Once the $S$ is discriminative, we stop adding landmark to it and roll back to previous level recursion.
Since the same $S$ may be generated in different order, to eliminate duplication, 
we only consider those landmarks with a lower significance than any element in $S$.
The process stops when all the possible combinations have been visited. 
\begin{algorithm}
\small
\caption{expansion}\label{algo:expansion}
\If{$|S|=n$}{
	stop or $S \gets $ landmark with the next biggest significance;
}
\Else{
	 $SetOfS \gets$ <- all the landmarks has a lower sinificance than any landmark of $S$;
	 
	 Sort $SetOfS$ in descending order of the significances of landmarks; 
	 
	 \For{each $l \in SetOfS$}{
	 	\If{$S \cup \{l\}$ is not discriminative}{
	 		expand($S\cup \{l\}$);
	 	}
	 }	 
}
\end{algorithm}

%

{\bf Test step:} this step tests whether $S$ is discriminative.
If $S$ is discriminative, calculate the maximum target value $m$ of $S$ and the supersets of $S$ and if $value$ is bigger than the current maximum target value $maxValue$, then the set with target value $value$ among $S$ and the super sets of $S$ will be assigned to $\mathbb{L}_{best}$.
Afterwards the supersets of $S$ do not to be visited since their are all discriminative and their maximum target value will be no more than $value$. 
Thus all the landmarks of $S$ will be removed, which in other words $S$ will be assigned as $\emptyset$.

However, the above two algorithms can be very time consuming when the size of $\mathbb{L}$ and $n$ are large, since there will be a large amount of landmark sets to be tested.
In order to improve the efficiency, we need to filter out more non-beneficial landmarks in the preparation step, generate $S$ more intelligently by avoiding duplication in the expansion step and test less $S$ which prunes many insignificant enough landmark sets and their supersets in the test step.
Next we will present the optimizations for each step respectively for both algorithms. 

\subsection{Question Ordering}

In the previous step we select questions (landmarks), which can be regarded as the {\em question library}.
However, presenting those question to workers with random order is unwise
because of the following two reasons:
1) it is not necessary to ask all the questions in most cases. For example, in Fig~\ref{fig:routes} if a worker indicate he prefers the routes passing $l_2$ from $l_1$ to $l_{10}$,  we do not need to ask whether he recommend to pass $l_8$ since all the routes passing $l_2$ do not pass $l_8$;
2) each time we ask a question, we would like to obtain the most informative feedback, which is more likely to indicate the final answer. 
This implies that 1) the next question to be asked depends on the result of the previous question, so the question order is a tree-like structure, denoted by $\mathcal{T}$;  
 2) the information strength of a question $q$ is proportional to people's familiarity of a landmark (the significance of the landmark) and the power of filtering of a landmark (the information gain of the candidate routes set using the landmark).

The information strength $IS(l_{k_i})$ of question $l_{k_i}$ is defined as following: 
{\footnotesize
\begin{align*}
IS(l_{k_i})= & l_{k_i}.s [H (\bar {\mathbb{R}}_{k-1}) - \textstyle{\bar {\mathbb{R}}_{k}^+ \over \bar {\mathbb{R}}_{k}^+ +\bar {\mathbb{R}}_{k}^-}H (\bar {\mathbb{R}}_{k}^+) -\textstyle{\bar {\mathbb{R}}_{k}^- \over \bar {\mathbb{R}}_{k}^+ +\bar {\mathbb{R}}_{k}^-}H (\bar {\mathbb{R}}_{k}^-) ]
\end{align*}}
where $H(\cdot)$ is the empirical entropy of $\cdot$, $\bar {\mathbb{R}}_{k-1}$ stands for  $\bar {\mathbb{R}}_{l_{k_1}^{+/-}l_{k_2}^{+/-}\cdots l_{k_{i-1}}^{+/-}}$ and $\bar {\mathbb{R}}_{k}^+$ and $\bar {\mathbb{R}}_{k}^-$ represent $\bar {\mathbb{R}}_{l_{k_1}^{+/-}\cdots l_{k_{i-1}}^{+/-}l_{k_i}^{+}}$ and $\bar {\mathbb{R}}_{l_{k_1}^{+/-}\cdots l_{k_{i-1}}^{+/-}l_{k_i}^{-}}$ respectively.

In order to get more information after each question, we employ the Iterative Dichotomiser 3 (ID3) algorithm~\cite{quinlan1986induction}, which recursively selects the question with the largest information strength as the next question, to build $\mathcal{T}$.
The algorithm can be mainly divided into four steps.
1) Calculate the information strength of every question using the whole routes set $\bar {\mathbb{R}}$.
2) Split the routes set $\bar {\mathbb{R}}_{l_{k_1}^{+/-}l_{k_2}^{+/-}\cdots l_{k_i}^{+/-}}$ into two subsets $\bar {\mathbb{R}}_{l_{k_1}^{+/-}\cdots l_{k_i}^{+/-}l_{k_{i+1}}^{+}}$ and $\bar {\mathbb{R}}_{l_{k_1}^{+/-}\cdots l_{k_i}^{+/-}l_{k_{i+1}}^{-}}$ according to the answer of question $k_{i+1}$, which has the maximum information strength among all questions.
Here we use $\bar {\mathbb{R}}_{l_{k_1}^{+/-}\cdots l_{k_i}^{+/-} l_{k_{i+1}}^{+/-}}$ to denote the routes subset after answering questions $l_{k_1},\cdots,l_{k_i}, l_{k_{i+1}}$ and the $k_{i+1}^+$ denotes that the answer of $k_{i+1}$ is yes and $k_{i+1}^-$ denotes that the answer of $k_{i+1}$ is no.
3) make a decision node of $\mathcal{T}$ containing question $k_{i+1}$.
4) perform the above steps recursively on routes subsets $\bar {\mathbb{R}}_{l_{k_1}^{+/-}\cdots l_{k_i}^{+/-}l_{k_{i+1}}^{+}}$ and $\bar {\mathbb{R}}_{l_{k_1}^{+/-}\cdots l_{k_i}^{+/-}l_{k_{i+1}}^{-}}$ using remaining questions until all the subsets have only one route.

\section{Worker Selection}\label{sec:targetworker}
Some Crowdsourcing platforms such as AMT and CrowdFlower give workers the freedom to choose any questions.
However this may cause some problems, for example, many workers choose to answer a same question while some other questions are not picked by anyone, workers have to view all the questions before they choose,  workers may answer questions that they are not familiar with. 
CrowdPlanner avoids these problems by  designing a dedicated component to assign each task to a set of eligible workers.
In order to judge whether a worker is eligible for a task, many aspects of the worker have to be taken into consideration, i.e., number of outstanding tasks, worker's response time and familiarity with a certain area.
First, since each worker may have many outstanding tasks, in order to balance the workload and reduce the response time, we use a threshold $\eta_{\#q}$ to restrict the maximum number of tasks of each worker.
Second, each user of CrowdPlanner can specify the longest time delay she allows to get an answer, so this task will not be assigned to workers who have high probability not to accomplish the task before the due time.
Last, a recommended route will have high confidence to be correct if assigned workers are very familiar with this area. 
Again, the worker's familiarity with respect to a certain area can also be affected by several factors, such as whether the worker lives around the area, whether the worker has answered questions relating to this area correctly in the past, etc. 
In summary, an eligible worker should meet three conditions: 1. has quota to answer the question; 2. has high probability to answer a question before due time;  3. has relatively high familiarity level with the query regions.


\subsection{Response Time} \label{set:Response Time}

Each task has a user-specified response time, with which an answer must be returned. 
We assume the probability of the response time $t$ of a worker follows an exponential distribution~\cite{Doe:2009:exponentialdistribution}, i.e., $f(t; \lambda) =\lambda \exp^{-\lambda t}$, which is standard assumption in estimating worker's response time.
The cumulative distribution function of $f(t;\lambda)$ is  $F(t;\lambda)= 1-\exp ^{-\lambda t}$. If the probability of a worker to respond a task within time $\overline t$, represented by $F (\overline t; \lambda)$, is less than the threshold $\eta_{time}$, we will not assign the task to him.

\subsection{Worker's Familiarity Score}

People usually have the best knowledge for areas where they live or travel about frequently. 
In CrowdPlanner, we develop a familiarity score $f_w^l$ to estimate the knowledge of a worker $w$ about a landmark $l$. 
$f_w^l$ is mainly affected by two factors: (1) worker's profile information, including her home address, work place and familiar suburbs, which can be collected during her registration to the system, and (2) history of worker's tasks around this area.
$f_w^l$ of landmark is defined as:

\begin{align*}
f_w^l =& \alpha\cdot\exp{\{
-(d(l,p_{home})+d(l,p_{work})+d(l, p_{fr})\}} \\
&+(1-\alpha)\cdot(\#correct+\beta\cdot \#wrong)
\end{align*}

where $\alpha$ is a smoothing variable, $d(l,p_{*})$ is the distance between $l$ and $p_{*}$, $\#correct$ is the number of correctly answered question of $l$, $\#wrong$ is the number of incorrectly answered question of $l$, and $\beta$ is a constant less than $1$, which measures the gain of a wrong answer.
Notably, since the knowledge of a far away region can hardly influence the knowledge here and in order to simplify the computation of $f_w^l$, we assign $+\infty$ to $d(l,x)$ if $d(l,p_{*})$ is bigger than a threshold $\eta_{dis}$. 
With all the $n$ workers and $m$ landmarks in our system, a $n*m$ matrix $M$ with $m_{ij}= f_{w_i}^{l_j}$ is built, where $f_{w_i}^{l_j}$ is worker $w_i$'s familiarity score of landmark $l_j$.
Since the number of landmarks a worker has answered is always small compared with the large number of landmarks in the space, $M$ is very sparse.
Hence, if task assigning is only based on the sparse $M$, the assigning process has a strong bias to assign tasks to only a few well-performed workers.
Actually, workers who have similar profile information or have answered several similar questions are highly possible to share the similar knowledge. For example, if a worker $w_1$ has high familiarity score with $l_1, l_2$ and $ l_3$ and another worker $w_2$ living nearby  has high familiarity score with $l_1$ and $l_2$, $w_2$ is also likely to be familiar with $l_3$ though $w_2$ has not answered any question relating to $l_3$.
Therefore, we need to predict familiarity scores of workers on landmarks using the latent similarity between workers.

The familiarity scores of different landmarks of workers are determined by some unweighed or even unobserved factors, which are regarded as some hidden knowledge categories, e.g. certain type of landmarks. However, we do not manually specify these factors, as hard-coded factors are usually limited and biased.
Instead, we assume the familiarity score of each worker-landmark pair is a linear combination of
two groups of scores, i.e. (1) how a worker is familiar with each hidden knowledge category, and (2) how a landmark
is related to each hidden knowledge category. Then we
employ Probabilistic Matrix Factorization (PMF)~\cite{mnih2007probabilistic} to factorize $M$ into two latent feature matrices, $W \in R^{d \times n}$ and $L^{d\times m}$, which are the latent worker and landmark feature matrices, respectively. 
That is,  $M = W^T L$,
where $W_{i,k}$ describes how familiar worker $w_i$ is with knowledge category $k$,
and $L_{j,l}$ describes how related landmark $l_j$ is to knowledge category $k$.
Further, we assume there exists observation uncertainty $R$, and the uncertain follows a normal distribution.
Thus the distribution of a new worker-landmark familiarity matrix $M'$, which predicts some familiarity by leveraging the similarity between different workers and landmarks, conditioned on $W$ and $L$ is defined as follows:

{\footnotesize
\begin{equation}
\label{equation:pmf1}
p(M'|W,L,{\sigma ^2}) = \prod\limits_{i = 1}^n {\prod\limits_{j = 1}^m {{{[\mathcal{N}({M_{ij}}|W_i^T{L_j},{\sigma ^2})]}^{{I_{ij}}}}} }
\end{equation}
}

where $\mathcal{N}(x|\mu, \theta^2)$ is the probability density function of the normal distribution with mean $\mu$ and variance $\theta^2$, and $I_{ij}$ is a indicator which is equal to 1 if $M_{ij}$ is not zero, otherwise 0.
The prior of $W$ and $L$ are defined as follows:

{
\begin{align*}
&p(W|\sigma_W^2)=\prod\limits_{i=1}^n \mathcal{N}(W_i|0,\sigma_W^2\boldmath{I} )\\ &p(L|\sigma_L^2)=\prod\limits_{i=1}^m \mathcal{N}(L_i|0,\sigma_L^2\boldmath{I} )
\end{align*}
}

where $\boldmath{I}$ is identity matrix.
The following objective function maximizes the posterior of $W$ and $L$ with regularization terms,
which minimizes the prediction difference between our model and the observed
$M$, and also automatically detects the appropriate number of factors $d$ through the regularization terms:

{\footnotesize
\begin{equation*}
\sum\limits_{i = 1}^n {\sum\limits_{j = 1}^m {{I_{ij}}{{({M_{ij}} - W_i^T{L_j})}^2} + } } {\lambda _W}\sum\limits_{i = 1}^n {\left\| {{W_i}} \right\|_F^2 + } {\lambda _L}\sum\limits_{j = 1}^m {\left\| {{L_j}} \right\|_F^2}
\end{equation*}
}

where $\lambda_W=\theta^2 / \theta_W^2$, $\lambda_L=\theta^2 / \theta_L^2$, and $\left\| {\cdot} \right\|^2_F$ denotes the Frobenius norm.
A local minimum of the objective function can be found by performing gradient descent in $W$ and $L$.
Afterwards, more familiarity scores between workers and landmarks are inferred in $M$.

A worker with a familiarity score of a landmark means he has some knowledge about the region around the landmark, not just the  landmark itself.
As a result, the accumulated familiarity score $F_{w_i}^{l_j}$ of $l_j$ of a worker $w_i$ is a weighted sum of all the landmarks in the $\eta_{dis}$ range of $l_j$.
We assume the weight around a landmark $l$ follows a normal distribution of the distance to $l$,
and the region that the knowledge of $l$ can cover is limited in a circle with the center of $l$ and the radius of $\eta_{dis} $.
Thus, $F_{w_i}^{l_j}$ is evaluated as follows:

{\footnotesize
\begin{equation*}
F_{w_i}^{l_j} = \sum\limits_{l \in \mathbb{L}_{near}  \cup \{l_j\}} {\delta_{l}{f_{w_i}^l}}
\end{equation*}
}

where $\mathbb{L}_{near}$ is the set of landmarks in the $\eta_{dis}$ range of $l$.
The weight $\delta_{l} = \mathcal{N}(d(l,l_j)|0, \sigma^2_0)$
where $\sigma_0 = \eta_{dis}/3$.
We use $M^*$ to denote the worker-landmark matrix of the accumulated familiarity score, where $m^*_{ij}$ equals to $ F_{w_i}^{l_j}$.

%

%

\subsection{Finding Top-k Eligible Workers}

Next we discuss how to find the top-k eligible workers for a given task.
Given a task (the selected $n$ landmarks $\mathbb{L}$), the worker-landmark accumulated familiarity score matrix $M^*$, a response time $t$, a positive integer $k$, a top-$k$ eligible workers query returns $k$ workers who have the most knowledge of landmarks in $\mathbb {L}$ among all the workers and have high possibility to finish the task within time $t$.

For a single landmark $l_j$, there may be several workers, denoted as $\mathbb{W}_{l_j}$, who have non-zero accumulated familiar scores, which means these workers have some knowledge of $l_j$.
For a task (a set of landmarks $\mathbb{L}$), $\bigcup \limits_{l\in \mathbb{L}}{\mathbb{W}_l}$ represents workers who have knowledge of any landmark of $\mathbb{L}$. 
Then we filter out workers, of who the possibility of finishing the task within time $t$ is no more than $\eta_{t}$, from $\bigcup \limits_{l\in \mathbb{L}}{\mathbb{W}_l}$. 
Afterwards the remained workers in $\bigcup \limits_{l\in \mathbb{L}}{\mathbb{W}_l}$ are regarded as candidate workers denoted by $\mathbb{W}$.
However, simply adding up a worker's accumulated familiarity scores on all the landmarks of $\mathbb{L}$ may lead biased result in worker selection.
For example, there are ten landmarks in a task and two candidates workers $w_1$ and $w_2$, that $w_1$ only has a very good knowledge of landmark $l_1$ , $F_{w_1}^{l_1}$=2) and knows nothing about the rest landmarks, $F_{w_1}^{l_i}=0$ ($2 \le i\le 10$), while $w_2$ has some knowledge of all the landmarks that $F_{w_2}(l_i)=0.1 $ ($1 \le i\le 10$).
Comparing the adding up sum of accumulated familiarity scores of the ten landmarks, $w_1$ will be selected to be assigned the task.
However, the coverage of $w_1$'s knowledge of the landmark set is too narrow, that $w_1$ may feel hard to answer questions about $l_2, l_3, \cdots, l_{10}$, in the knowledge coverage manner, $w_2$ is a better choice.
Thus, when selecting workers from candidate workers, not only their sum of accumulated familiarity scores of all the landmarks, but also the knowledge coverage of all the landmarks should be considered.
The choosing rules are quite similar to rated voting system~\cite{Doe:2009:Votingsystem}, of which the wining option is chosen according to the voters preferences score of options and the number of voters preferring the options. 
In our system, we can treat each landmark of $\mathbb{L}$ as a voter and each worker of $\mathbb{W}$ as an option.
Adopting the idea of rated voting system, we can measure the landmark $l_j$'s preference of all the candidate workers as the following two steps: 1) rank workers of $\mathbb{W}_{l_j} \cap \mathbb{W}$, who  are in the candidate workers set $\mathbb{W}$ and have accumulated familiar scores $F_w^{l_j}$ bigger than zero, in descending order of $F_w^{l_j}$; 2) the preference score $p_{l_j}(w)$ of  $l_j$ to each worker $w$ in $\mathbb{W}_{l_j} \cap \mathbb{W}$ is defined as follows:

{\footnotesize
\begin{equation*}
p_{l_j}(w) = \left\{ \begin{array}{l}
 1 - \frac{{rank(w) - 1}}{{|W_{l_j}|}},if~w\in W_{l_j} \\
 0,otherwise \\
 \end{array} \right.
\end{equation*}
}

where $rank(w)$ is the ranked place of $w$ among $\mathbb{W}_{l_j} \cap \mathbb{W}$.
In this way the worker with high accumulate familiarity score will get a relatively high preference score and ensure the preference score will not result in a bias in worker selecting.
Afterwards, all the landmarks will vote their preferences to the candidate workers, at this time we sum up the preferences of each worker voted by landmarks and then the workers with the top-$k$ biggest adding up preference scores will be the query results.

\section{Related Work}\label{sec:related}
To our knowledge, there is no existing work on evaluating the quality of recommended routes. 
As the goal of this work is to evaluate the quality of recommended routes by web services and mining algorithms, the route recommendation algorithms (mining frequent path algorithms) used in this paper are reviewed first. 
Then we will review route generating algorithms.
Since in our Crowdplanner system, we reuse truth to reduce the request times of Crowdsourcing, which share similar inspiration and techniques with route generating.
Also we leverage the generating easy questions and  finding target workers to improve the quality of evaluating and reduce the workers' workload, which share the same motivation of some research works of Crowdsourcing question designing and workers selecting.  
Therefore in the last of this section, we will review these two lines of related work. 

{\bf Route Recommendation Algorithms.}
The popular routes mining has received tremendous research interests for a decade and a lot of works are on it, such as \cite{sacharidis2008line, mamoulis2004mining, giannotti2007trajectory, gonzalez2007adaptive, zheng2009mining,giannotti2007trajectory, mamoulis2004mining,gonzalez2007adaptive,zheng2009mining,li2007traffic,lee2007trajectory,gaffney1999trajectory,lee2008traclass}.
Among these works, \cite{chen2011discovering, wei2012constructing,Luo2013Finding, Ceikut2013Routing} are the most representative.   
Chen et al. \cite{chen2011discovering} proposes a novel popularity function for
path desirability evaluation using historical
trajectory datasets. 
The popular routes recommended by it tends to have fewer vertices. 
The work in~\cite{wei2012constructing} provides k popular routes by mining uncertain trajectories. 
The recommendation routes of this work tend to be rough routes instead of correct routes. 
\cite{Luo2013Finding} claims the popular routes change as time, so it carries out a popular routes mining algorithms which can provide the recommended routes in arbitrary time periods specified by the users. 
\cite{Ceikut2013Routing} provides the evidences that the routes recommended by web services are sometimes different from drivers' preference. 
Thus it mines the individual popular routes from his historical trajectories. 
The recommended routes of this method reflect certain people's preference. 

{\bf Question Designing.}
Question designing is always an application dependent strategy, which may consider the cost of questions or the number of questions. 
\cite{guo2012so, parameswaran2012crowdscreen} propose strategies to minimize the cost of the questions designed.
The question designing strategy of~\cite{wang2012crowder} is to minimize the number of questions. 
The question designing strategy of~\cite{parameswaran2011human} is to generate the optimal set of questions. 

{\bf Worker Selecting.}
Selecting workers with high individual qualities for tasks always does beneficial to the final quality of answers. 
Thus~\cite{lappas2009finding} propose an algorithm to select workers fulfilling some skills with the minimized the cost of choosing them. 
In~\cite{campbell2003expertise} use emails communication to identifying skillful workers. 
Cao et al~\cite{cao2012whom} assign tasks to micro-blog users by mining users' knowledge and measuring their error rate.

\section{Conclusions}\label{sec:conclusion}

In this paper we have taken an important step towards evaluating recommended routes from different providers, web services and popular routes mining algorithms, to provide users the verified best routes. 
After studying the difference between web service recommended routes and popular routes mined from historical trajectories, we have proposed a system CrowdPlanner which give users the verified best routes and allow computers together with crowds to evaluate routes effectively and efficiently. 
We use many components to reduce the time cost and manpower cost of the route evaluating procedure, such as,  truth reusing, landmark selecting and worker selecting, where in this paper we details the human evaluating related part. 
Landmark selecting automatically generates an identifiable set of landmarks with the highest mean significance. 
Worker selecting finds the top-k most eligible workers who have good knowledge of the task. 
Extensive experiments have been conducted involving a lot of volunteers and using a real trajectory dataset. 
We have demonstrated that the CrowdPlanner system can always give users the best routes. 
The MFP (Mining Frequent Path) has the highest possibility to give the best route. 
The ideas from this work open a new direction for future research, such as quality control of popular route mining algorithms, and mining latent factor which may affect drivers' driving routes.

\bibliographystyle{abbrv}
\bibliography{myRef}
\end{document}